# BAB 4
# *Web Based Learning*

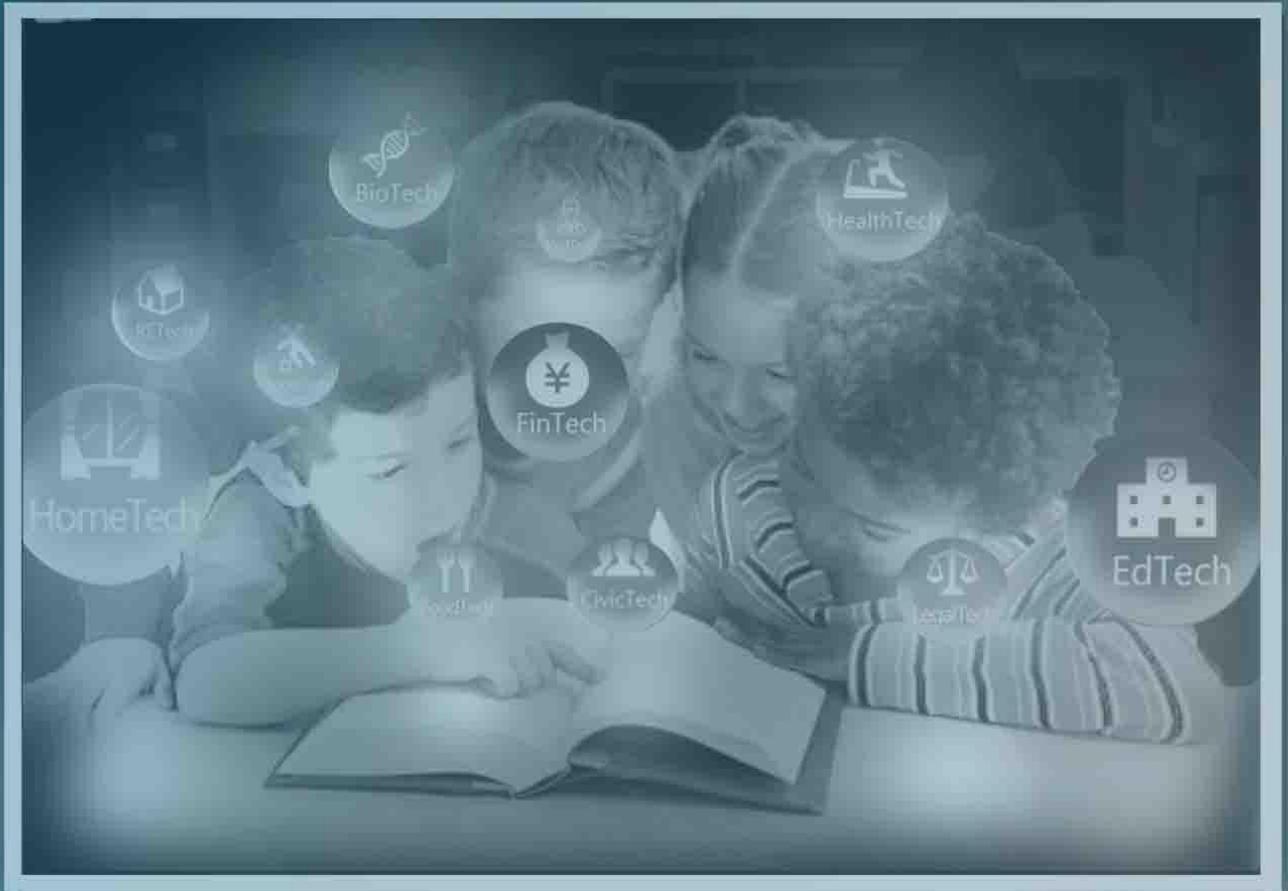

# Leon A. Abdillah

# BAB 4
# *WEB-BASED LEARNING*

### A. *Web-based Learning Overview*

Teknologi Informasi (TI) telah memberikan sejumlah dampak positif di berbagai bidang (Abdillah *et al.*, 2020). Pada dunia bidang pendidikan, TI memberikan corak tersendiri dengan berbagai mode yang mungkin digunakan. Pendidikan modern di masa globalisasi dan berbasis Teknologi Informasi telah bertransformasi ke arah *digital*. TI telah menjadi tulang punggung proses pembelajaran modern (Abdillah *et al.*, 2021). Proses pendidikan yang tadinya dilakukan secara klasik dengan mode tatap muka secara langsung telah mengalami pergeseran ke arah mode pembelajaran jarak jauh (*distance learning*). Pada mode pembelajaran *distance learning*, para peserta didik dapat mengakses materi perkuliahan melalui internet.

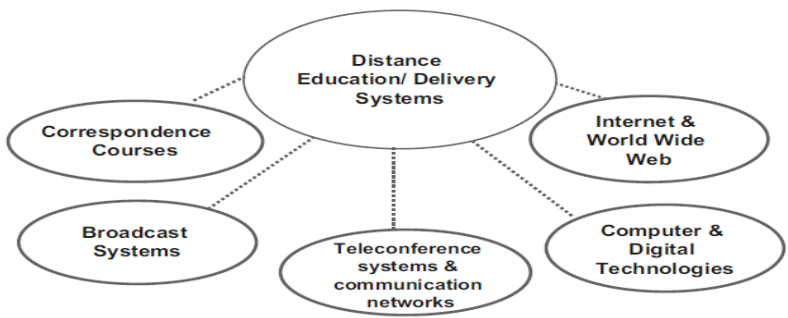

Gambar 4.1 Tipe Sistem *Distance Education* (Davidson-Shivers, Rasmussen and Lowenthal, 2018)

Internet dan *World Wide Web* atau Web menciptakan sejumlah pilihan pendidikan jarak jauh (*Distance Education*). Setidaknya ada 5 (lima) macam bentuk *Distance Education* (Davidson-Shivers, Rasmussen and Lowenthal, 2018) seperti yang nampak pada Gambar 4.1, yaitu:
1. *Internet & World Wide Web*,
2. *Computer & Digital Technologies*,
3. *Teleconference Systems & Communication Networks*,
4. *Broadcast Systems*, dan
5. *Correspondence Courses*.

Khusus untuk pembelajaran yang memberdayakan internet dan WWW atau dikenal dengan *e-learning* memiliki 5 (lima) tipe (Horton and Horton, 2003), yaitu:
1. *Learner-led e-learning*.
2. *Facilitated e-learning*.
3. *Instructor-led e-learning*.
4. *Embedded e-learning*.
5. *Telementoring and e-coaching*.

Pada bab ini, penulis akan mengulas "Web-Based Learning" sebagai salah satu trend TI di bidang pendidikan. Pembelajaran berbasis *web* sering juga disebut *online learning* atau *e-learning* karena di dalamnya terdapat konten kursus *online* (McKimm, Jollie and Cantillon, 2003) dan mencakup semua intervensi pendidikan yang menggunakan internet (atau intranet lokal) (Cook, 2007).

Selanjutnya bab ini akan membahas:
1. Web-based Learning Framework
2. Advantages & Disadvantages Web-based Learning Model

   3. Developing Web-based Learning
   4. Web-based Learning Menggunakan Platform Moodle
   5. Virtual Web-based Learning Menggunakan WordPress
   6. Social Media Web-based Learning
   7. Web-based Learning Evaluation by using Google Forms

**B. *Web-based Learning Framework***

*Web-Based Learning* (WBL) mengacu pada penggunaan teknologi Internet untuk menyampaikan instruksi (Aggarwal, 2003) dari dosen/guru/pendidik ke mahasiswa/siswa/peserta. Ketika merancang suatu *Web-Based Learning* perlu memerhatikan 8 (delapan) dimensi dari *Web-Based Learning* atau *E-Learning Framework* (Khan, 2005):

   1. *Pedagogical*
   2. *Technological*
   3. *Interface Design*
   4. *Evaluation*
   5. *Management*
   6. *Resource Support*
   7. *Ethical*
   8. *Institutional*

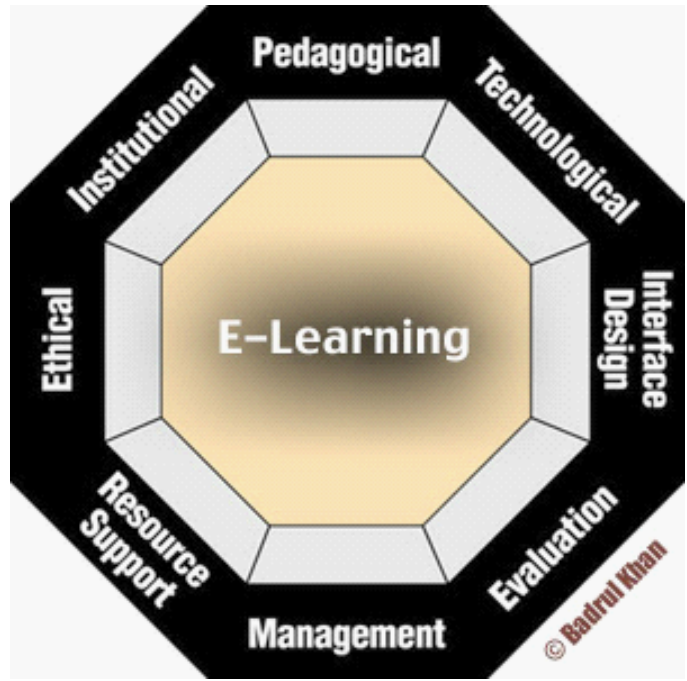

**Gambar 4.2**
*Web-Based Learning Framework* **(Khan, 2005)**

Lebih lanjut, *Web-Based Learning* atau *E-Learning Framework* yang ada di Gambar 4.2 dapat diuraikan dalam bentuk deskripsi yang nampak pada Tabel 4.1.

**Tabel 4.1 Dimensi *E-Learning***

| Dimensi E-Learning | Deskripsi |
|---|---|
| *Institutional* | Dimensi kelembagaan berkaitan dengan masalah administrasi, akademik, dan layanan kemahasiswaan yang berkaitan |

| Dimensi E-Learning | Deskripsi |
|---|---|
| | dengan *e-learning*. |
| *Management* | Manajemen *e-learning* mengacu pada pemeliharaan lingkungan belajar dan distribusi informasi. |
| *Technological* | Dimensi teknologi e-learning mengkaji masalah infrastruktur teknologi di lingkungan *e-learning*. Ini termasuk perencanaan infrastruktur, perangkat keras, dan perangkat lunak. |
| *Pedagogical* | Dimensi pedagogis *e-learning* mengacu pada pengajaran dan pembelajaran. Dimensi ini membahas masalah-masalah yang berkaitan dengan analisis konten, analisis audiens, analisis tujuan, analisis media, pendekatan desain, organisasi, dan strategi pembelajaran. |
| *Ethical* | Pertimbangan etis dari *e-learning* berkaitan dengan pengaruh sosial dan politik, keragaman budaya, bias, keragaman geografis, keragaman pelajar, kesenjangan digital, etiket, dan masalah hukum. |
| *Interface design* | Desain antarmuka mengacu pada tampilan dan nuansa program *e-learning* secara keseluruhan. Dimensi desain antarmuka mencakup desain halaman dan situs, desain konten, navigasi, aksesibilitas, dan pengujian kegunaan. |
| *Resource support* | Dimensi dukungan sumber daya dari *e-learning* memeriksa dukungan online dan sumber daya yang diperlukan untuk |

| Dimensi E-Learning | Deskripsi |
|---|---|
|  | mendorong pembelajaran yang bermakna. |
| *Evaluation* | Evaluasi untuk *e-learning* mencakup penilaian peserta didik dan evaluasi instruksi dan lingkungan belajar. |

Kedelapan dimensi tersebut akan memudahkan para pengembang WBL ketika membangun suatu WBL, baik yang bersifat *open source* maupun yang dibangun dari awal.

### C. *Advantages & Disadvantages Web-based Learning Model*

WBL menawarkan *Virtual Learning Environment* (VLE) atau *Internet-based Training* (IBT) yang komprehensif baik untuk para pendidik maupun peserta didik. Dengan VLE yang berbasiskan TI, memberikan pengalaman baru dana berbeda. Berikut ada sejumlah keuntungan dan kerugian dari penerapan WBL.

Adapun keuntungan (*advantages*) dari model *Web-Based Learning* (McKimm, Jollie and Cantillon, 2003; Cook, 2007), sebagai berikut:

1. Kemampuan untuk menghubungkan (*link*) sumber daya dalam berbagai format.
2. Dapat menjadi cara yang efisien (*efficient*) dalam menyampaikan materi kursus, termasuk *distance learning*.
3. Sumber daya dapat tersedia (*available*) dari lokasi mana pun dan kapan pun. Sumber daya abadi (*perpetual resource*) yang mudah diperbarui.

4. Potensi untuk memperluas (*wide*) akses — misalnya, ke siswa paruh waktu, dewasa, atau berbasis pekerjaan. WBL menawarkan penjadwalan yanag fleksibel (*flexible scheduling*).
5. Dapat mendorong pembelajaran lebih mandiri dan aktif (*independent & active learning*), serta mendukung pembelajaran individual (*Individualised learning*).
6. Dapat memberikan sumber bahan pelengkap (*supplementary*) yang berguna untuk program konvensional.
7. Metode pembelajaran baru (*novel instructional methods*), baik pada penympaian materi maupun pelaksanaan asesmen dan dokumentasi (*assessment and documentation*).

Selain sejumlah keuntungan diatas, penerapan WBL juga memiliki sejulah kerugian (*disadvantages*), sebagai berikut (McKimm, Jollie and Cantillon, 2003; Cook, 2007):

1. Akses ke peralatan komputer yang sesuai dapat menjadi masalah (*problem*) bagi siswa, yaitu berupa masalah teknis (*technical problems*).
2. Peserta didik merasa frustasi (*frustating*) jika mereka tidak dapat mengakses grafik, gambar, dan klip video karena peralatan yang buruk.
3. Infrastruktur yang diperlukan harus tersedia dan terjangkau. Hal ini bisa menilnulkan biaya (*cost*) tambahan.
4. Kualitas dan keakuratan informasi dapat berbeda-beda, sehingga diperlukan panduan dan penunjuk arah. Jika tidak jelas, maka terjadi Instruksi de-individualisasi (*de-individualised instruction*).

5. Siswa dapat merasa terisolasi secara sosial (*social isolation*).
6. Desain instruksional yang buruk (*poor instructional design*) dapat membuat WBL menjadi tidak efektif.

## D. Developing Web-based Learning

Untuk meningkatkan proses belajar mengajar sebaiknya memerhatikan 3 (tiga) area dasar (*fundamental*) yang meliputi (Song and Kidd, 2009):
1. Desain (*Design*): Identifikasi sistematis kebutuhan pendidikan dan desain, pengembangan, implementasi dan evaluasi bahan untuk digunakan di ruang kelas dan di tempat yang jauh.
2. Pengembangan (*Development*): Produksi materi untuk memenuhi tujuan pendidikan tertentu termasuk multimedia program, grafik, dan video.
3. Penelitian/Evaluasi (*Research/Evaluation*): Perencanaan, desain, dan pelaksanaan penelitian dan/atau pengembangan proyek yang menerapkan prinsip-prinsip teknologi pendidikan untuk setiap aspek pendidikan atau pelatihan dalam berbagai pengaturan, termasuk sekolah, industri, kedokteran, dan militer.

Sedangkan tahapan pengembangan *website* pendidikan terdiri dari 3 (tiga) tahap (Astuti, Wihardi and Rochintaniawati, 2020), yaitu:
1. Tahapan analisis (*Analysis*).
   a. Analisis material (*material*).
   b. Analisis pengguna (*user*).

  c. Analisis kebutuhan perangkat lunak (*software*).
  d. Analisis kebutuhan perangkat keras (*hardware*).
2. Tahapan desain (*Design*).
  a. Desain materi pembelajaran (*learning material*).
  b. Desain diagram alur (*flowchart*).
  c. Desain papan cerita (*storyboard*).
3. Tahap pengembangan (*Development*).
  a. Pembangunan antarmuka (*interface*).
  b. Pembangunan pembuatan koding (*coding*).

### E. *Web-based Learning Menggunakan Platform Moodle*

  Moodle adalah salah satu *Learning Management Systems* (LMS) paling populer untuk model *Web-Based Learning*. Moodle mungkin paling populer di kalangan perguruan tinggi dan universitas karena Moodle merupakan *platform* berbasis pengembang kaya fitur yang cocok untuk perguruan tinggi dan universitas dengan anggaran terbatas (Fenton, 2018).

  Moodle tidak hanya sebagai salah satu varian WBL atau *e-learning*, namun Moodile juga bisa dimanfaatkan sebagai fasilitator *Knowledge Management Systems* (KMS) di bidang pendidikan. Sejumlah fasilitas yang disediakan oleh Moodle memungkinnya untuk menjadi media KMS yang cukup handal. Moodle bisa digunakan untuk *Information and Knowledge Sharing* dari sumber ke destinasi. Sumber dapat berupa orang atau individu atau kelompok dalam suatu organisasi. Destinasi juga bisa serupa dengan sumber tetapi mereka mungkin dari organisasi yang sama atau dari luar organisasi tertentu (Abdillah, 2014).

Tampilan awal Moodle *dashboard* nampak seperti Gambar 4.3. Pada bagian "Recently accessed courses" nempak sejumlah mata kuliah yang baru saja diakses. Pada contoh ini ada 4 (empat) mata kuliah yang baru saja diakses, yaitu:
1. *Business Process in Procurement*.
2. *Analysis and Design Systems*.
3. *Human-Computer Interaction*.
4. *Management at Administration*.

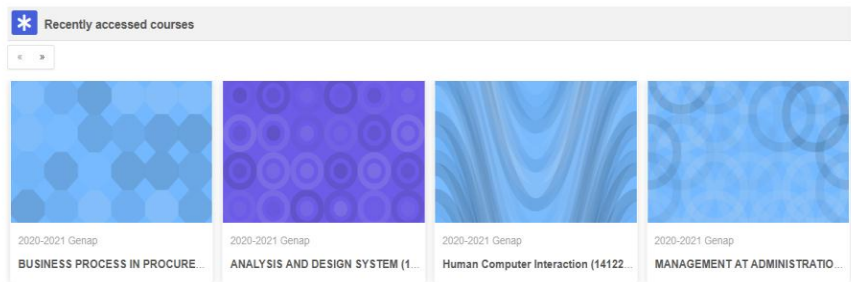

**Gambar 4.3 Tampilan** *Moodle Dashboard*

Moodle menyediakan sejumlah *features* yang dapat digunakan untuk berinteraksi dengan para peserta didik. Fasilitas tersebut dibagi menjadi 2 (dua) kelompok (Tabel 4.2), yaitu:
1. *Activities*.
2. *Resources*.

**Tabel 4.2 Fasilitas Moodle**

| *Activities* | *Resources* |
|---|---|
| 1) *Assignment* | 1) *Book* |
| 2) *Chat* | 2) *edu-sharing resource* |
| 3) *Choice* | 3) *File* |
| 4) *Custom certificate* | 4) *Folder* |
| 5) *Database* | 5) *IMS content package* |
| 6) *External tool* | 6) *Label* |
| 7) *Feedback* | 7) *Page* |
| 8) *Forum* | 8) *URL* |
| 9) *Glossary* | |
| 10) *Lesson* | |
| 11) *Quiz* | |
| 12) *SCORM package* | |
| 13) *Survey* | |
| 14) *Wiki* | |
| 15) *Workshop* | |

Untuk contoh pada mata kuliah "Analysis and Design Systems" pendidik dapat mengubah periode belajar permingggunya menjadi "01) 15 February – 22 February". Artinya, pertemuan pertama dihitung mulai dari tanggal 15 – 22 Februari. Kemudian dengan menggunakan fasilitas "Label" dapat menampilkan "Label" untuk pertemuan pertama berupa "01) INTRODUCTION".

ZOOM Meetings dilibatkan sebagai media tatap muka secara *online*. Selanjutnya *screen shoot* dari ZOOM Meetings ditampilkan sebagai bukti bahwa telah berlangusng perkuliahan secara *online* pada pertemuan 01. Jika diperlukan, Moodle juga memiliki fasilitas *Unified Resource Locator* (URL) untuk memasukkan *link* materi dari sumber-sumber lainnya.

Interaksi dengan para peserta didik juga bisa memanfaatkan fasilitas "Forum". Sedangkan materi perkuliahan dibuat dalam bentuk *Portabel Document Format* (PDF) yang disematkan melalui fasilitas "File".

Gambar 4.4 menampilkan contoh pertemuan pada Moodle yang dikombinasikan dengan aplikasi ZOOM Meetings, sehingga perkuliahan dapat berjalan dengan mode "Blended Learning" juga. ZOOM Meetings (L. A. Abdillah, 2020) digunakan untuk melakukan pertemuan tatap muka secara *online* dengan para peserta didik (*synchronous*). Fasilitas "Forum" dimanfaatkan untuk menerima respon dari para peserta didik secara *asynchronous*.

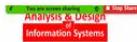

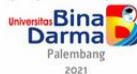
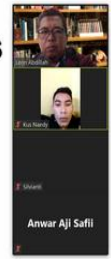

**Gambar 4.4 Pertemuan di *Moodle***

## F.　Virtual Web-based Learning Menggunakan WordPress

Varian lain dari WBL ada dengan memanfaatkan *blog platforms* seperti WordPress. *Open-source technologies* seperti WordPress membawa serta akses ke komunitas dukungan yang komprehensif, termasuk pembaruan gratis, teknologi terkini, dan berbagai alat untuk penyesuaian yang dibuat dan ditawarkan (kebanyakan) secara bebas oleh komunitas (Roseth, Akcaoglu and Zellner, 2013). Dengan menggunakan WordPress, para pendidik dapat dengan mudah dan fleksibel menampilkan materi-materi kurssu yang akan disampaikan kepada para peserta didik.

Pada contoh kali ini, *blog* dengan menggunakan WordPress dimanfaatkan untuk menampilkan daftar materi per mata kuliah yang tidak hanya bisa diakses oleh para peserta didik, namun juga bisa diakses oleh *public*. WordPress memiliki fasilitas berupa "Page" untuk mengelompokkan sejumlah "Post" dari masing-masing materi kursus.

Pada contoh yang nampak seperti pada Gambar 4.5, daftar kursus atau mata kuliah untuk satu semester di-list berupa label mata kuliahnya. Label tersebut berisikan hyperlink ke halaman "Page" dari masing-masing mata kuliah.

**Gambar 4.5 Daftar Mata Kuliah di WordPress**

Setiap mata kuliah akan berisikan informasi seputar konten dari mata kuliah tersebut. Salah satu cara untuk menarik mahasiswa misalnya dengan menampilkan *overview* dari mata kuliah yang bersangkutan dengan menggunakan infografis (Gambar 4.6). di infografis tersebut berisikan: *Descriptions, Prerequesite Courses, Assesments, Lengths, Tools,* dan *References*.

Gambar 4.6 Contoh Mata Kuliah A&DIS di WordPress

## G. Social Media Web-based Learning

Web 2.0 membuka sejumlah aplikasi yang selanjutnya menjadi trend di dunia internet. *Social Media* muncul sebagai salah satu *instance* aplikasi berbasiskan internet yang banyak diminati oleh para penggunanya. Media sosial sebagai salah satu aplikasi internet menawarkan banyak manfaat bagi sistem pembelajaran modern. TI telah mengubah cara gaya dan pendekatan belajar. Media sosial dapat digunakan untuk mendukung TI atau *cyber universities* (Abdillah, 2016).

*"Model Pembelajaran Era Society 5.0"*

Berbagai *platforms* bermunculan dengan keunggulan dan ciri khasnya masing-masing. Pada sub-bab ini akan diulas pemanfataan Facebook dalam pembelajaran berbasiskan *web*.

Facebook telah menjelma menjadi rajanya *social media*, bahkan menjadi salah satu *icon* di dunia TI yang memiliki sangat banyak penggunanya. Facebook adalah jejaring sosial pertama yang melampaui 1 (satu) miliar akun terdaftar (L. Abdillah, 2020). Walaupun Facebook tidak secara khusus dibuat untuk kepentingan pendidikan, namun Facebook memiliki sejumlah fasilitas yang bisa diatur menjadi *semi-vitual learning environment*. Fasilitas yang bisa digunakan untuk membuat *virtual class* adalah "Group". Pendidik dapat mengatur *Facebook Group* (FBG) menjadi *private*. Para peserta didik harus *join* untuk bisa masuk ke FBG tersebut melalui URL-nya.

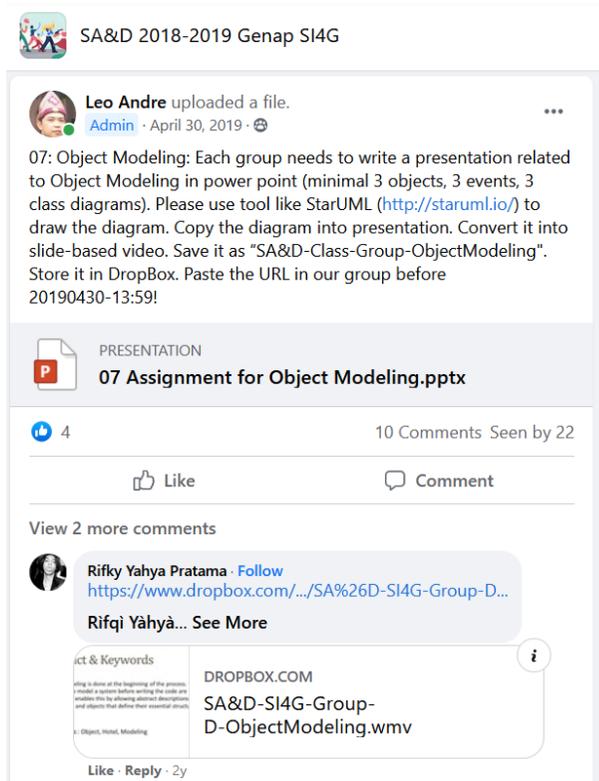

Gambar 4.7 Contoh *Assignment* Mata Kuliah A&DIS di Facebook

Pada contoh Gambar 4.7, FBG dimanfaatkan untuk menerima pengumpulan tugas mahasiswa. Para mahasiswa mengerjakan tugas kelompok yang disimpan di "DropBox". DropBox digunakan untuk menyimpan laporan kelompok siswa (Abdillah, Sari and Indriani, 2018). Kemudian, URL dari DropBox dikumpulkan di FBG mata kuliah yang bersangkutan.

"Model Pembelajaran Era *Society* 5.0"

## H. Web-based Learning Evaluation by using Google Forms

Salah satu kegiatan terpenting pada suatu perkuliahan adalah aktivitas evaluasi atau ujian. Evaluasi bisa berupa Ujian Harian (UH), Ujian Tengah Semester (UTS), atau Ujian Akhir Semester (UAS). Nilai atau skor yang didapat dari ujian-ujian tersebut dapat digunakan oleh para pendidik untuk memetakan capaian para peserta didik dalam bentuk nilai (angka atau huruf).

Pada mode *Web-Based Learning Evaluation* menggunakan "Google Forms", para pendidik dapat membuat skema evaluasi secara *online* baik untuk skema ujian *synchronous* maupun *asynchrinous* (Gambar 4.8). Pada ujian menggunakan Google Forms, para pendidik dapat menggunakan sejumlah pilihan format untuk soalnya. Google Forms menyediakan pilihan feedback untuk jawaban pertanyaan-pertanyyan dalah bentuk:

1. *Multiple choice*.
2. *Short answer*.
3. *Paragraph*.
4. *Checkboxes*.
5. *Drop-down*.
6. *Linear scale*.
7. *Multiple-choice grid*.
8. *Tick box grid*.
9. *File upload*.
10. *Date*.
11. *Time*.

Gambar 4.8 Contoh Soal Ujian A&DIS Menggunakan Google Forms

Setiap soal juga bisa diberi bobot, sehingga setelah para peserta didik *submit* ujiannya, Google Forms dapat secara otomatis menghitung total skor yang didapat. Fasilitas ini sangat membantu proses pemberian nilai atau *marking*. Google Forms juga dilengkapi dengan fasilitas yang memungkinkan rekapitulasi dalambentuk bagan-bagan sehingga memudahkan proses analisis evaluasi hasil ujian para peserta didik.

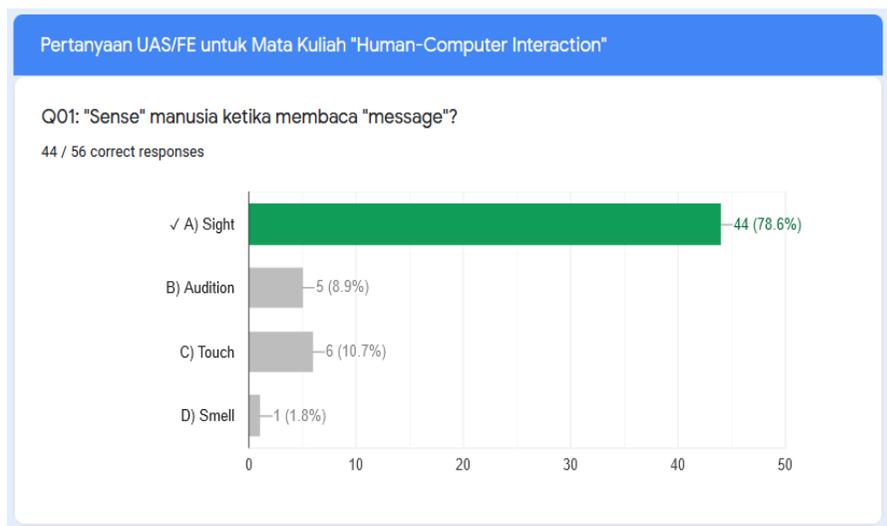

Gambar 4.9 Bagan Rekapitulasi Jawaban Benar dengan Google Forms

# DAFTAR PUSTAKA

## PROFIL PENULIS

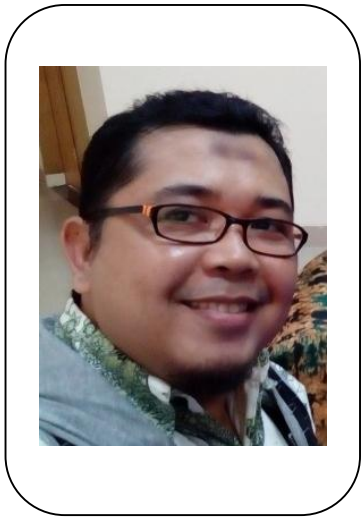

**Leon A. Abdillah,** lahir di Limau Barat, Prabumulih/Muara Enim, Sumatera Selatan. Ia pernah belajar *Information Systems*, *Information Systems Management*, dan *Information Retrieval Systems* selama masa studi. Tahun 2001 bergabung sebagai dosen di salah satu perguruan tinggi swasta terkemuka di Kota Palembang. Tahun 2010 menjadi *Associate Professor* (Assoc. Prof.) pada Fakultas Ilmu Komputer Program Studi Sistem Informasi (Terakreditasi A). Assoc. Prof. Leon A. Abdillah aktif menjadi *speaker*, *author*, *editor*, *reviewer*, *committee* pada sejumlah *journals*, *conferences/seminars*, *books/book chapters*, dll. Beliau termasuk 500 Indonesian scientist (Webometrics, 2015), *examiner* di Monash University (*Group of Eight*), Australia, dan *mentor* di Publons, New Zealand. Beliau juga sering mendapatkan *awards* untuk kategori *best undergraduate and post graduate*, *the best computer science lecturer*, *the best reference article*, *excellent paper*, *top reviewer, selected article*, dsb.